\documentclass[10pt]{iopart}
\usepackage{iopams}
\expandafter\let\csname equation*\endcsname\relax
\expandafter\let\csname endequation*\endcsname\relax
\usepackage{amsmath}
\usepackage{graphicx,bm}
\usepackage{xcolor}
\usepackage{multirow}

\begin{document}
\title{Computational study of III-V direct-gap semiconductors for thermoradiative cell applications}
\author{Muhammad~Yusrul~Hanna$^{1, 2, \dagger}$, Muhammad~Aziz~Majidi$^{1, \ddag}$,
Ahmad~R.~T.~Nugraha$^{2,3,4,*}$}
\address{$^{1}$ Department of Physics, Faculty of Mathematics and Natural Sciences, Universitas Indonesia, Depok 16424, Indonesia}
\address{$^{2}$ Research Center for Quantum Physics, National Research and Innovation Agency (BRIN), South Tangerang 15314, Indonesia}
\address{$^{3}$ Research Collaboration Center for Quantum Technology 2.0, Bandung 40132, Indonesia}
\address{$^{4}$ Department of Engineering Physics, Telkom University, Bandung 40257, Indonesia}
\ead{$^{\dagger}$muhammad.yusrul@ui.ac.id,
$^{\ddag}$aziz.majidi@sci.ui.ac.id,
$^{*}$ahmad.ridwan.tresna.nugraha@brin.go.id}
\begin{abstract}
We investigate the performance of thermoradiative (TR) cells using the III-V group of semiconductors, which include GaAs, GaSb, InAs, and InP, with the aim of determining their efficiency and finding the best TR cell materials among the III-V group.  The TR cells generate electricity from thermal radiation, and their efficiency is influenced by several factors such as the bandgap, temperature difference, and absorption spectrum.  To create a realistic model, we incorporate sub-bandgap and heat losses in our calculations and utilize density-functional theory to determine the energy gap and optical properties of each material.  Our findings suggest that the effect of absorptivity on the material, especially when the sub-bandgap and heat losses are considered, can decrease the efficiency of TR cells.  However, careful treatment of the absorptivity indicates that not all materials have the same trend of decrease in the TR cell efficiency when taking the loss mechanisms into account.  We observe that GaSb exhibits the highest power density, while InP demonstrates the lowest one.  Moreover, GaAs and InP exhibit relatively high efficiency without the sub-bandgap and heat losses, whereas InAs display lower efficiency without considering the losses, yet exhibit higher resistance to sub-bandgap and heat losses compared to the other materials, thus effectively becoming the best TR cell material in the III-V group of semiconductors.
\end{abstract}
\submitto{\NT}
%
\maketitle
%
\ioptwocol
%
\section{Introduction}
Clean and sustainable energy conversion technology, especially utilizing ``free" resources, such as sun and waste heat, is in high demand to solve various environmental and energy issues.  Recently, there has been a growing interest in using negative illuminated energy converters to generate electric power.  These devices maintain a temperature higher than their surroundings and require a constant heat flow to operate.  Not only does the input heat sustain the device's temperature, but it also generates the output power.  In 2014, Byrnes et~al.~\cite{Byrnes2014} proposed some methods to obtain electric power from Earth's thermal infrared emission.  One of the methods is by using optoelectronic harvesters.   The optoelectronic harvesters are similar to thermophotovoltaic cells and are easy to implement, hence referred to as thermoradiative (TR) cells~\cite{Strandberg2015,Santhanam2016}.  

\begin{figure}[tb]
        \centering
        \includegraphics[clip, width=7cm]{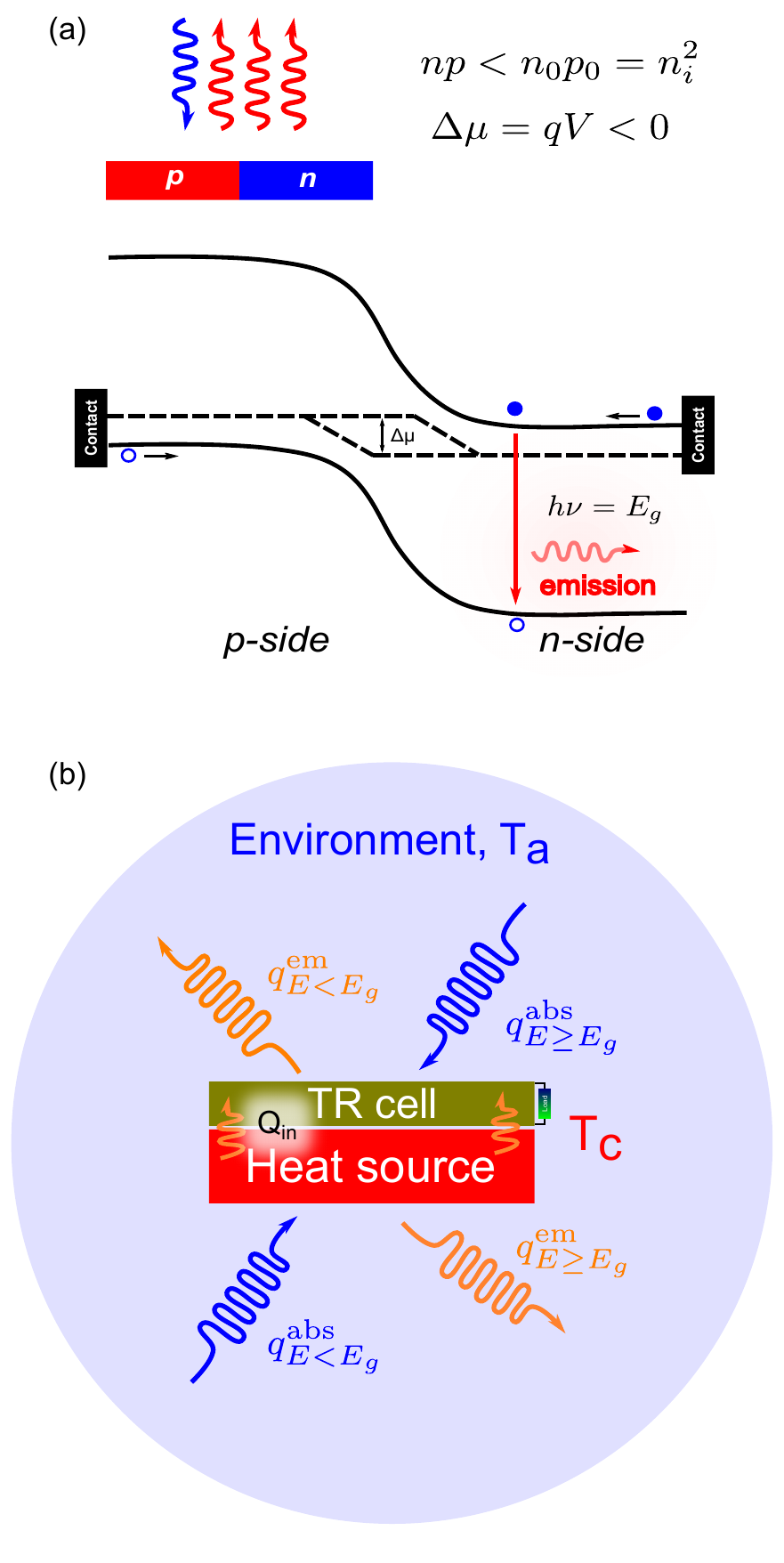}
        \caption{(a) The schematic diagram and (b) the energy conversion process of TR cells to generate electricity from the emission of thermal radiation.}
        \label{fig:TRcellSchematic}
    \end{figure}

A TR cell consists of a $p$-$n$ junction, which could generate electricity from the thermal radiation through electron-hole pair generations from external photons~\cite{Strandberg2015,Hsu2016,Zhang2017,Deppe2020}.  When the $p$-$n$ junction of the TR cell operates at a higher temperature than its surroundings (see figure~\ref{fig:TRcellSchematic}(a)), the device emits more energy than it absorbs.   On the other hand, if the TR cells are connected to a thermal reservoir, the temperature is maintained at a constant level.  The excess emission of energy decreases the carrier concentration, resulting in a splitting of the electron and hole Fermi levels in the opposite direction and introducing a reverse bias voltage across the junction.  Under short-circuit or load-connected conditions, the loss of an electron-hole pair that is not balanced by photon absorption causes an additional electron and hole to be injected via the contacts to restore the lost electron-hole pair, thus leading to current flow~\cite{Hsu2016,Deppe2020}.  However, the inclusion of loss mechanisms such as sub-bandgap and heat losses leads to a substantial reduction in the TR cells efficiency~\cite{Strandberg2015,Hsu2016,Zhang2019,Feng2022}.

The sub-bandgap loss arises when photons with energies lower than the bandgap energy are emitted via blackbody radiation between the TR cell and the environment, while the heat loss occurs when radiative energy is released to the surroundings and heat from the environment is absorbed through conduction.  Previous research conducted by Strandberg~\cite{Strandberg2015} in 2015  investigated the impact of sub-bandgap loss on InAs material performance in TR cells within the detailed-balance model without careful treatment of optical absorption.  At a cell temperature of $T_c=500$ K and bandgap energy of $0.26$ eV, an ideal TR cell showed an efficiency of approximately $34\%$, which decreased to $\sim5\%$ when considering the sub-bandgap loss.  Hsu~et~al.~\cite{Hsu2016} and Zhang~et~al.~\cite{Zhang2019} also reported a significant reduction in efficiency due to sub-bandgap loss in narrow-bandgap InSb materials, with Zhang~et~al.~\cite{Zhang2019} investigating heat loss as well.  One possible way to reduce the sub-bandgap loss is by nanostructuring bulk materials.  For example, Feng~et~al.~\cite{Feng2022} proposed a solid-state waste heat recovery technology using an InSb-hBN TR cell with a nanoscale vacuum gap of $10$ nm, achieving efficiencies up to $39.6\%$.  Despite the sub-bandgap loss, the maximum efficiency of this device remained at an appreciable $30.6\%$.  Therefore, to obtain the most efficient TR cells, we need to search for bulk materials having a minimal sub-bandgap loss with ease of nano-structurization.  In this regard, the group III-V semiconductors may serve as a testbed. 

 Recent studies have focused on mitigating sub-bandgap losses in the III-V group material to improve the TR cell efficiency.  The III-V group of materials, such as GaAs, GaSb, InAs, and InP, have high electron mobility~\cite{Rode1970} and low thermal conductivity~\cite{Morelli2002}, making them ideal for TR applications.  This is due to their ability to transport electrons efficiently and reduce heat loss~\cite{Lin2020}.  Nanometer-scale transistors based on these materials, such as GaAs, InAs, and InP, have been used in many high-speed and high-frequency electronic systems~\cite{Alamo2011,Ajayan2015}.  Additionally, core-shell InAs/InP nanowires have been found to have superior electrical~\cite{Goeransson2019} and optical characteristics~\cite{Zhao2021}, core-shell InAs/InP nanowires exhibit superior electrical and optical characteristics, indicating that the in-situ grown InP layer can effectively suppress carrier leakage from the InAs nanowire core~\cite{Tilburg2010}.  Group III-antimonides such as GaSb~\cite{Borg2017} and (In)GaSb~\cite{Nagaiah2011,Kim2021} also have high electron and hole mobility.  For example, Hall/van der Pauw measurements conducted on $20$ nm-thick GaSb nanostructures revealed a high hole mobility of $760 \, \mathrm{cm}^2/(\mathrm{V s})$~\cite{Borg2017}, which matches literature values for high-quality bulk GaSb~\cite{Mikhailova1996}, making them a viable option for TR applications.

In this work, we examine the efficiency of TR cells considering the sub-bandgap and heat losses with a careful treatment of optical absorption, unlike Strandberg who neglected the absorption term in his 2015's work~\cite{Strandberg2015}.  We focus on semiconducting materials from the III-V group that have direct bandgaps, represented by GaAs, GaSb, InAs, and InP.  These materials exhibit high electron mobility and low thermal conductivity so they are expected to transport electrons efficiently and reduce heat loss~\cite{Lin2020}.  Furthermore, the III-V group forms the zinc-blende structure which means their band structure exhibits parabolic bands that can affect the TR cell efficiency by determining the number and energy of electrons participating in radiative transitions~\cite{Ye2018,Fernandez2022}.  The efficiency of a TR cell may depend on several factors such as the band gaps, temperature difference, emissivity, and absorption spectrum~\cite{Fernandez2022}.  Therefore, to determine the performance of TR cells, we require the energy gap and optical properties of each material using density-functional theory (DFT) calculations.  From our result, the effect of absorptivity as a function of photon energy on the material, particularly when sub-bandgap and heat losses are introduced, can decrease the efficiency of TR cells when compared to the constant absorptivity assumption used in the detailed-balance model.  Among the four materials considered in this study, we found that the power density of GaSb is the highest among the four materials considered, while InP exhibits the lowest power density.  GaAs and InP also demonstrate relatively high efficiency when the sub-bandgap loss is not considered, while InAs possess lower efficiency.  However, InAs demonstrate a relatively high efficiency when the sub-bandgap and heat losses is considered, in contrast to Strandberg's results, indicating that this material is more resistant to loss compared to the other materials.  

\section{Computational Methods}
\begin{figure}[tb]
        \centering
        \includegraphics[clip, width=8cm]{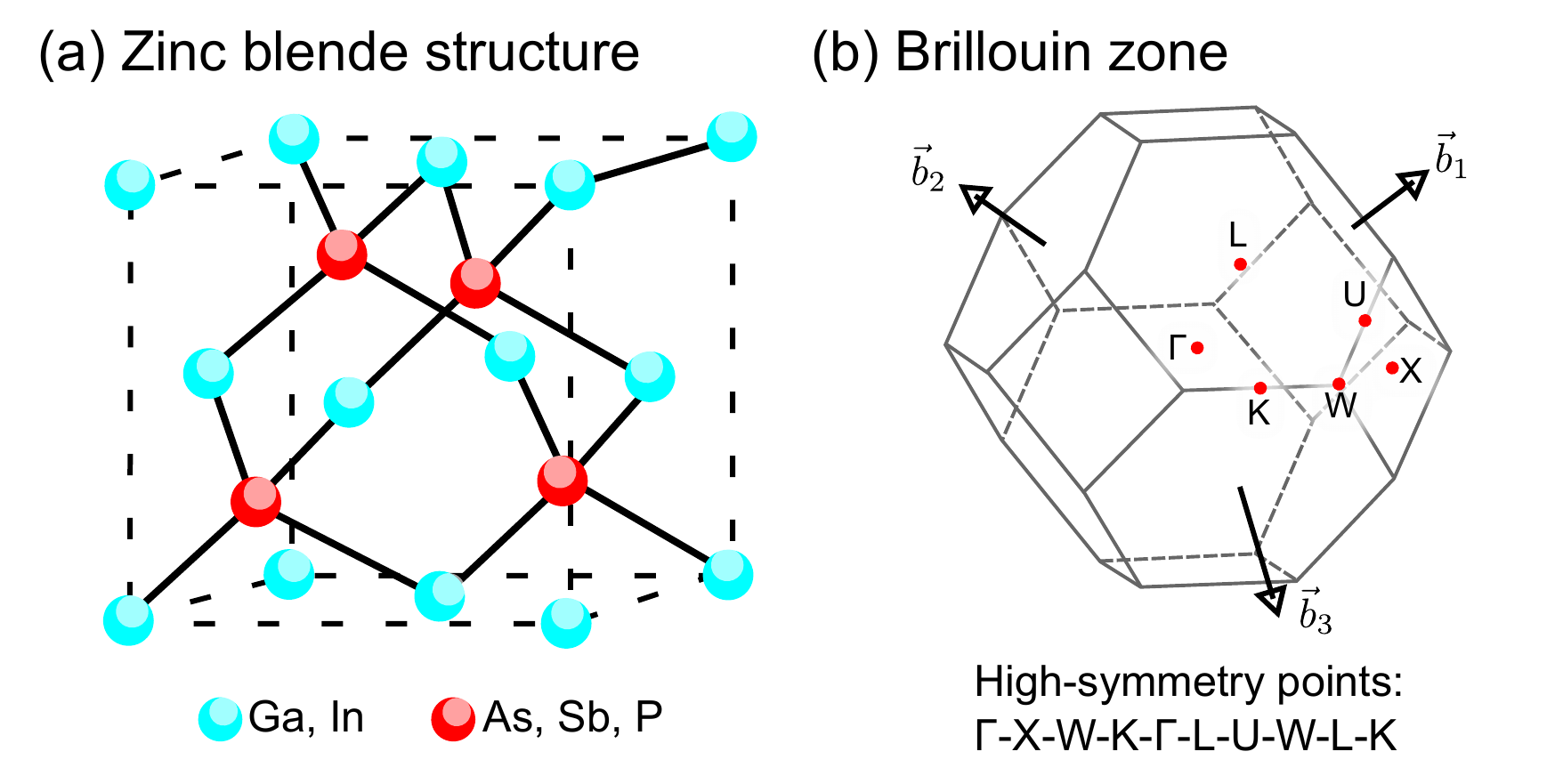}
        \caption{(a) Structure geometry of the zinc-blende, and (b) Brillouin zone in the reciprocal space.}
        \label{fig:structure}
    \end{figure}

To obtain the ground-state electronic structures of the materials, as the first important step before proceeding to other kinds of calculations, we employ the first-principles DFT calculations with a plane-wave basis, as implemented in the {\sc{Quantum ESPRESSO}} package~\cite{QE-2017}.  The ground-state electronic structure is calculated with a norm-conserving pseudopotential and the Perdew-Zunger (PZ)~\cite{Perdew1981} local density approximation (LDA)~\cite{SM} for the exchange-correlation functional along with a kinetic energy cutoff of $80$ Ry.  For this study, the LDA without spin-orbit interaction is used.  To perform self-consistent-field calculations, an $8 \times 8 \times 8$ Monkhorst-Pack k-mesh is employed to sample the Brillouin zone integration.  The system is represented as a cubic lattice with a zinc-blende structure (see figure~\ref{fig:structure}(a)). 

Having information on the bandgap energy and also optical properties, we could calculate the TR cell performance represented by power density and efficiency.  In the presence of an external load, the thermally excited electrons and holes are able to move through the $n$-type and $p$-type regions, respectively, and contribute to the current density ($J$) in the TR cell.  Therefore, the current density is formulated by~\cite{Strandberg2015}
\begin{equation}
        J = q \left[ N^{\mathrm{abs}}_{~}(T_{a}, \Delta \mu = 0) - N^{\mathrm{em}}_{~}(T_{c}, \Delta \mu) \right],
    \label{eq:1}
\end{equation}
where $q$ is the elementary charge and $T_{a}$ and $T_{c}$ represent the ambient and cell temperatures, respectively.  Here, $N^{\mathrm{abs}}_{~}$ denotes the photon flux absorbed from the environment, while $N^{\mathrm{em}}_{~}$ refers to the photon flux emitted by the cell into the environment.  Then, the formula for the photon flux density is given by~\cite{Ruppel1980,Wurfel2005}
\begin{equation}
        N(T, \Delta \mu, \hbar \omega) = \frac{1}{\pi^2 c^{2}_{0} \hbar^3}  \int^{\infty}_{\hbar \omega_{g}} \frac{a(\hbar \omega) n^2(\hbar \omega) (\hbar \omega)^2}{\left[\exp{\left(\frac{\hbar \omega - \Delta \mu}{k_B T}\right)-1}\right]} d(\hbar \omega)
    \label{eq:2}
\end{equation}
where $k_{B}$ and $\Delta \mu$ are the Boltzmann constant and the chemical potential, respectively. Our model assumes that the absorptivity has multiple internal reflections, given by $a(\hbar \omega) = [1 - R(\hbar \omega)][1 - \exp(-\alpha(\hbar \omega)d)]$. Here, $R(\hbar \omega)$ is the reflectivity, $\alpha (\hbar \omega)$ is the absorption coefficient of the material, and $n(\hbar \omega)$ is the normal refractive index.  The first law of thermodynamics is satisfied by the conversion of part of heat energy into electricity, depicted in figure~\ref{fig:TRcellSchematic}(b).   Hence, the energy balance of the TR cell can be expressed as follows~\cite{Zhang2017}:
\begin{equation}
        Q_{\mathrm{in}} + q^{\mathrm{abs}}_{E \geq E_g} + q^{\mathrm{abs}}_{E < E_g} = P + q^{\mathrm{em}}_{E \geq E_g} + q^{\mathrm{em}}_{E < E_g}
    \label{eq:3}
\end{equation}

In this work, we presume the TR cell to have a loss mechanism, including sub-bandgap losses, to create a more realistic model.  Thus, it is necessary to establish the radiative heat flux density for emission and absorption processes, both below and above the bandgap energy.  The radiative heat flux density absorbed by the TR cell, below and above the bandgap, is respectively expressed as~\cite{Zhang2017,Wurfel2005}:
\begin{align}
        \begin{aligned}
        q^{\text{abs}}_{\hbar \omega< \hbar \omega_{g}} &= \frac{1}{\pi^2 c^{2}_{0} \hbar^3}  \int^{\hbar \omega_{g}}_{0} \frac{a(\hbar \omega) n^2(\hbar \omega) (\hbar \omega)^3}{\left[\exp{\left(\frac{\hbar \omega}{k_B T_{a}}\right)-1}\right]} d(\hbar \omega) \\
        q^{\text{abs}}_{\hbar \omega \geq \hbar \omega_{g}} &= \frac{1}{\pi^2 c^{2}_{0} \hbar^3}  \int^{\infty}_{\hbar \omega_{g}} \frac{a(\hbar \omega) n^2(\hbar \omega) (\hbar \omega)^3}{\left[\exp{\left(\frac{\hbar \omega}{k_B T_{a}}\right)-1}\right]} d(\hbar \omega)
        \end{aligned}
    \label{eq:4}
\end{align}
Moreover, the expression for the radiative heat flux density below and above the bandgap emitted by the TR cell is given by~\cite{Zhang2017,Wurfel2005}
\begin{align}
        \begin{aligned}
            q^{\text{em}}_{\hbar \omega< \hbar \omega_{g}} &= \frac{1}{\pi^2 c^{2}_{0} \hbar^3}  \int^{\hbar \omega_{g}}_{0} \frac{a(\hbar \omega) n^2(\hbar \omega) (\hbar \omega)^3}{\left[\exp{\left(\frac{\hbar \omega}{k_B T_{c}}\right)-1}\right]} d(\hbar \omega) \\
            q^{\text{em}}_{\hbar \omega \geq \hbar \omega_{g}} &= \frac{1}{\pi^2 c^{2}_{0} \hbar^3}  \int^{\infty}_{\hbar \omega_{g}} \frac{a(\hbar \omega) n^2(\hbar \omega) (\hbar \omega)^3}{\left[\exp{\left(\frac{\hbar \omega - \Delta \mu}{k_B T_{c}}\right)-1}\right]} d(\hbar \omega)
        \end{aligned}
    \label{eq:5}
\end{align}
The integrations in the previous equations (see equations~\eqref{eq:2},~\eqref{eq:4} and~\eqref{eq:5}) are carried out from the bandgap energy $\hbar \omega_{g}$ to infinity.  The resulting photon and radiative heat flux densities are obtained from these integrations.  The photon energy range considered is from $0.0$ to $15$ eV, which includes visible light, ultraviolet light, and some parts of the infrared spectrum.  The study of material's absorption and emission of radiation within this energy range is of great importance. It should be noted that TR cells generate both power and heat by releasing radiative energy to their surroundings while simultaneously absorbing heat from the environment via conduction. As a result, there are extra losses in $Q_{in}$ in equation~\eqref{eq:3} represented by $q^{~}_{C} = U (T_c - T_a)$, where $U$ is the global heat transfer coefficient for the TR cell. We use a $U = 2 \, \text{W} \text{m}^{-2} \text{K}^{-1}$~\cite{Zhang2019} to indicate the effectiveness of heat transfer between the cell and its surroundings.

According to equations~\eqref{eq:1}--\eqref{eq:5}, the power density can be expressed as  
\begin{equation}
        P = \Delta \mu \left[ N^{\mathrm{abs}}_{~}(T_{a}, \Delta \mu = 0) - N^{\mathrm{em}}_{~}(T_{c}, \Delta \mu) \right],
\label{eq:6}
\end{equation}
and the efficiency of the TR cell including the sub-bandgap loss as the main contribution to the loss mechanism can be expressed as 
\begin{align}
        \eta &= \frac{P}{Q_{\text{in}}} \nonumber \\
        &= \frac{P}{P + q^{\text{em}}_{\hbar \omega< \hbar \omega_{g}} + q^{\text{em}}_{\hbar \omega \geq \hbar \omega_{g}} + q^{~}_{C}- q^{\text{abs}}_{\hbar \omega< \hbar \omega_{g}} - q^{\text{abs}}_{\hbar \omega \geq \hbar \omega_{g}}}. 
\end{align}
Note that, if generated in a material, the defect states such as vacancies, interstitials, or impurities could indirectly contribute to the sub-bandgap loss by introducing localized energy levels within the band gap, which can participate in radiative transitions.  In our calculations, since we consider perfect crystals of GaAs, GaSb, InP, and InAs, the defect states should be absent and are not relevant to the sub-bandgap losses in our results.  For ease of reproducibility, we also provide all the detailed calculations performed in this study in a GitHub~\cite{github} repository in form of Python codes with Jupyter notebooks.  

    \begin{figure}[tb]
        \centering
        \includegraphics[clip, width=8cm]{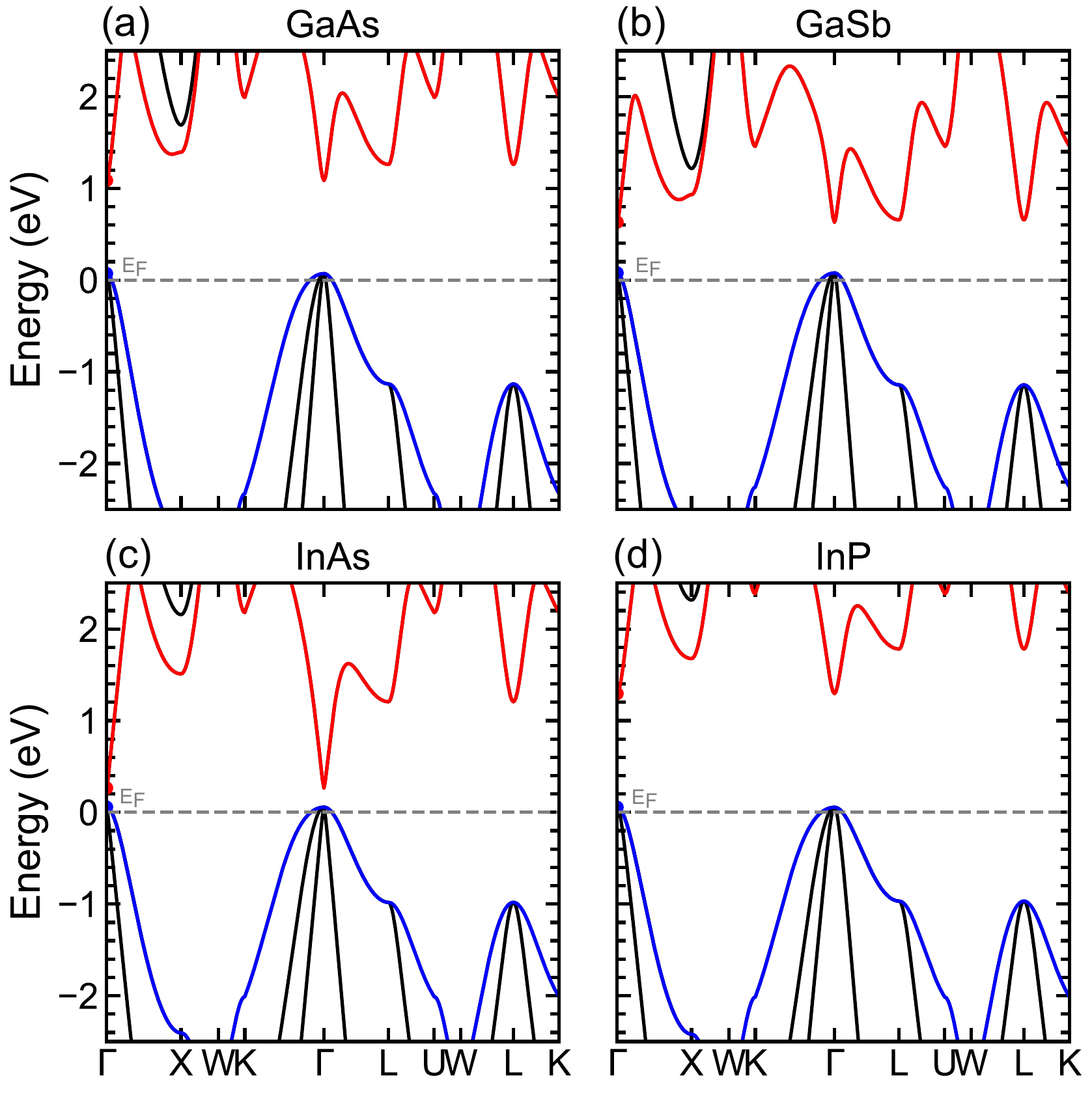}
        \caption{Band structures of (a) GaAs, (b) GaSb, (c) InAs, and (d) InP.  The band structures are plotted along the high-symmetry point direction in the Brillouin zone (following figure~\ref{fig:structure}(b)) and the Fermi level is indicated by the horizontal dashed line.  All band structures show the direct bandgaps.}
        \label{fig:bandstructure}
    \end{figure}
 
\section{Results and Discussion}

     \begin{figure}[tb]
            \centering
            \includegraphics[width=8cm]{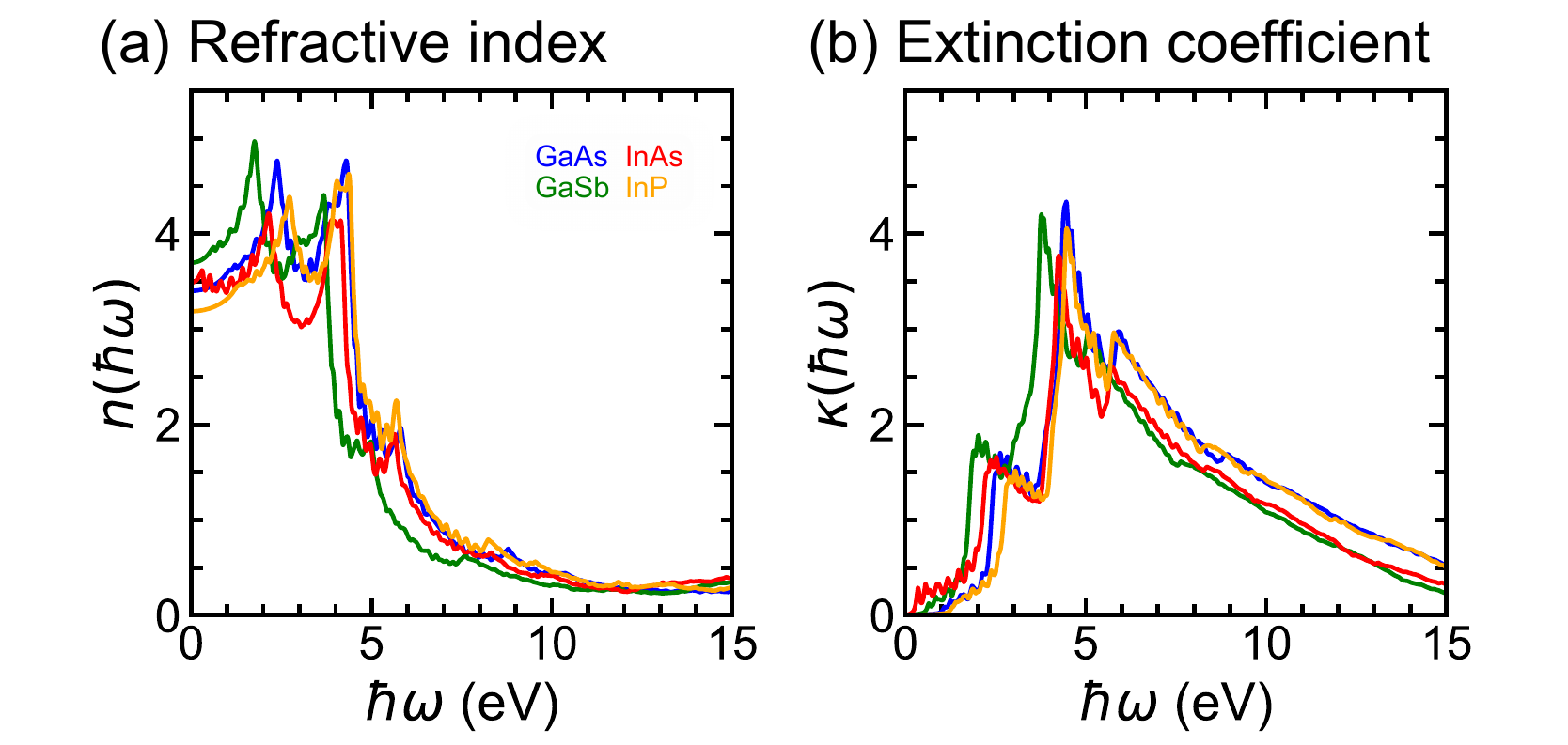}
            \caption{(a) Refractive index and (b) extinction coefficient as a function of photon energy for GaAs (blue solid line), GaSb (green solid line), InAs (red solid line), and InP (orange solid line).}
            \label{fig:optical}
        \end{figure}
        
The electronic band structures of bulk GaAs, GaSb, InAs, and InP are presented in figure~\ref{fig:bandstructure}, where the energy dispersion is plotted along specific high-symmetry points in the Brillouin zone of each material following the convention of figure~\ref{fig:structure}(b).  It can be seen that all of the materials exhibit semiconducting properties with direct band gaps.  The bandgap value for each material is provided in Table~\ref{tab:1}.      
        
    \begin{table}[tb]
        \centering
        \begin{tabular}{ccc}
        \hline
        {\textbf{Material}} & {\textbf{This work (eV)}} &{\textbf{Ref.~\cite{Mikhailova1996} (eV)}} \\ \hline
        GaAs & $1.016$ & $1.42$ \\
        GaSb & $0.554$ & $0.73$ \\
        InAs & $0.209$ & $0.36$ \\
        InP  & $1.240$ & $1.35$ \\ \hline
        \end{tabular}
    \caption{The bandgap energies (in eV) of GaAs, GaSb, InAs, and InP obtained in this work and from reference data.}
    \label{tab:1}
    \end{table}  
    
    \begin{figure}[tb]
        \centering
        \includegraphics[clip, width=8cm]{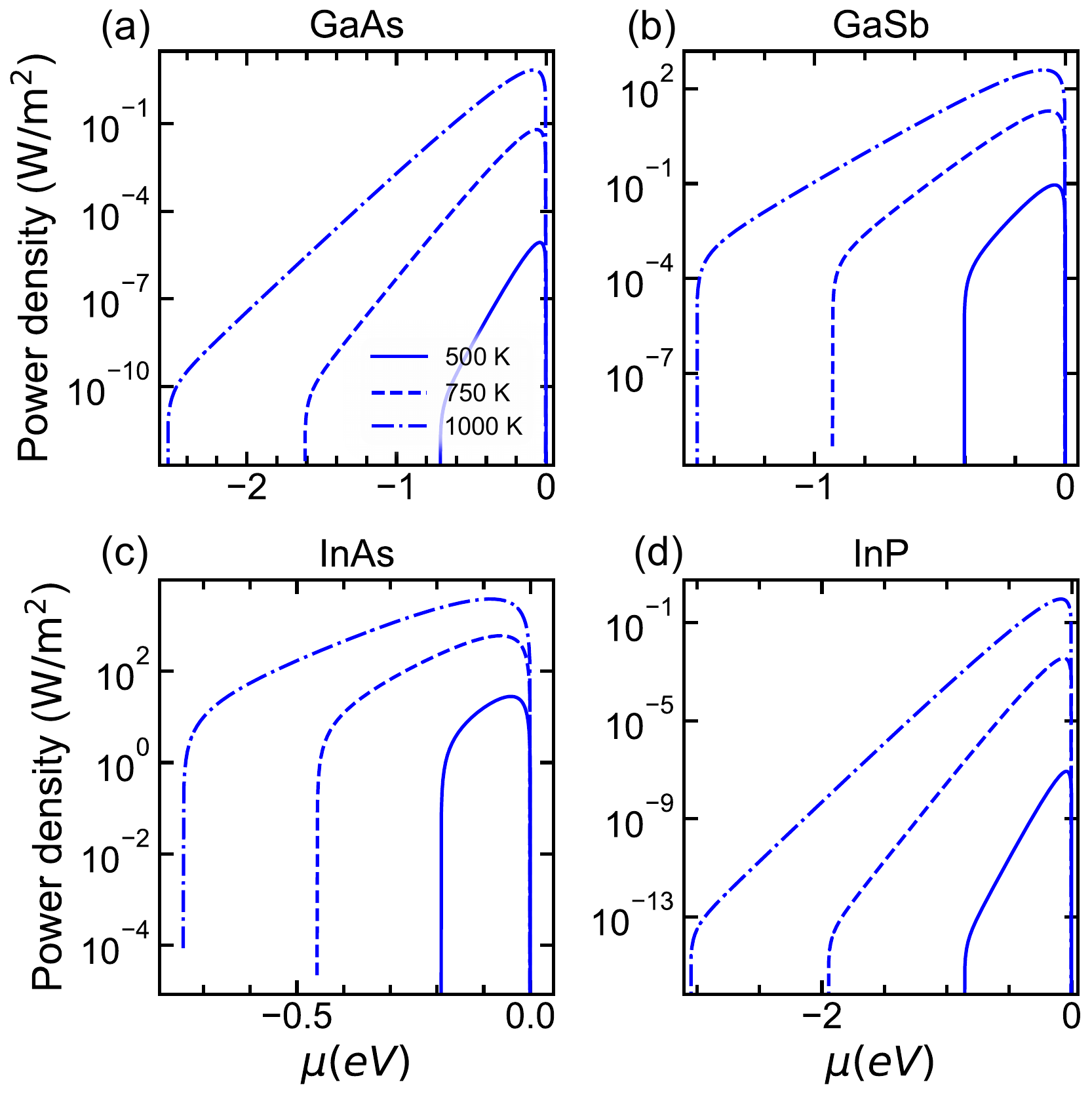}
        \caption{The power density of logarithmic plots is shown as a function of chemical potential for TR cells based on the following materials: (a) GaAs, (b) GaSb, (c) InAs, and (d) InP. The calculations were taken at various cell temperatures ($500$, $750$, and $1000$ K) and an ambient temperature of $300$ K.}
        \label{fig:power-density}
    \end{figure}
    
According to the results presented in Table~\ref{tab:1}, the bandgap energies obtained in this study are in good agreement with those reported in the reference experimental data, possibly due to our use of norm-conserving pseudopotentials with the LDA functionals.  This agreement is particularly significant as the bandgap energy is a crucial factor in determining the performance of the TR cell.  It is worth noting that the DFT calculations are performed at 0K in the ground states so that the calculated band gaps do not significantly depend on temperature.  Furthermore, as long as we ignore the interactions between electrons, electron-phonon, or electrons with other quasi-particles, the band structures from DFT calculations are practically temperature-independent as well~\cite{Allen1976, Wang2022}.  However, the TR cell performance indeed depends on the cell temperature as we will discuss later.

    \begin{figure}[tb]
        \centering
        \includegraphics[clip,width=0.8\linewidth]{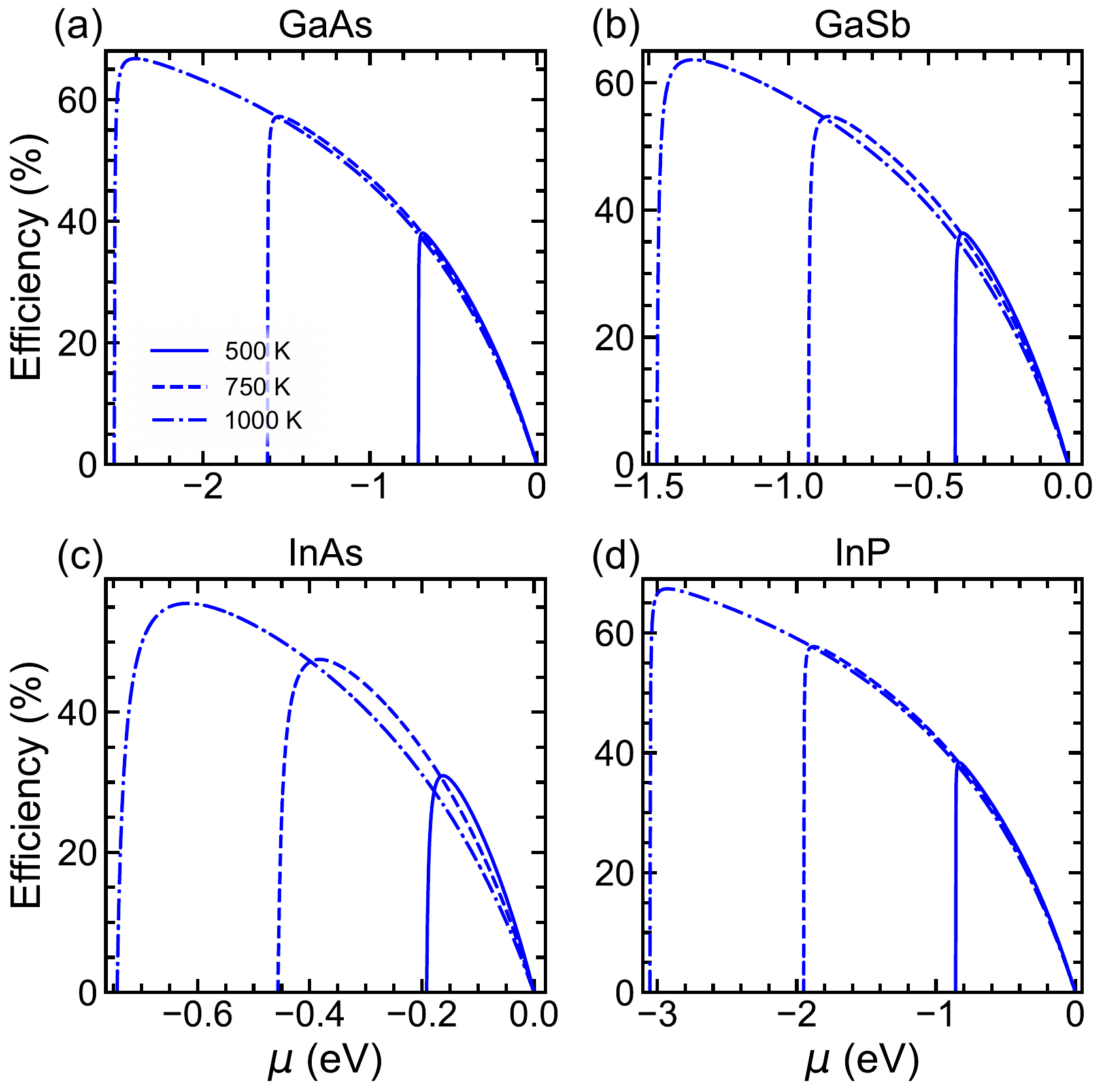}
        \caption{The efficiency of TR cells is shown as a function of chemical potential for different materials, including (a) GaAs, (b) GaSb, (c) InAs, and (d) InP. These calculations were performed at various cell temperatures of $500$, $750$, and $1000$ K, as well as at an ambient temperature of $300$ K, without accounting for any losses.}
        \label{fig:efficiencywolosses}
    \end{figure}

    \begin{figure}[tb]
        \centering
        \includegraphics[clip,width=8cm]{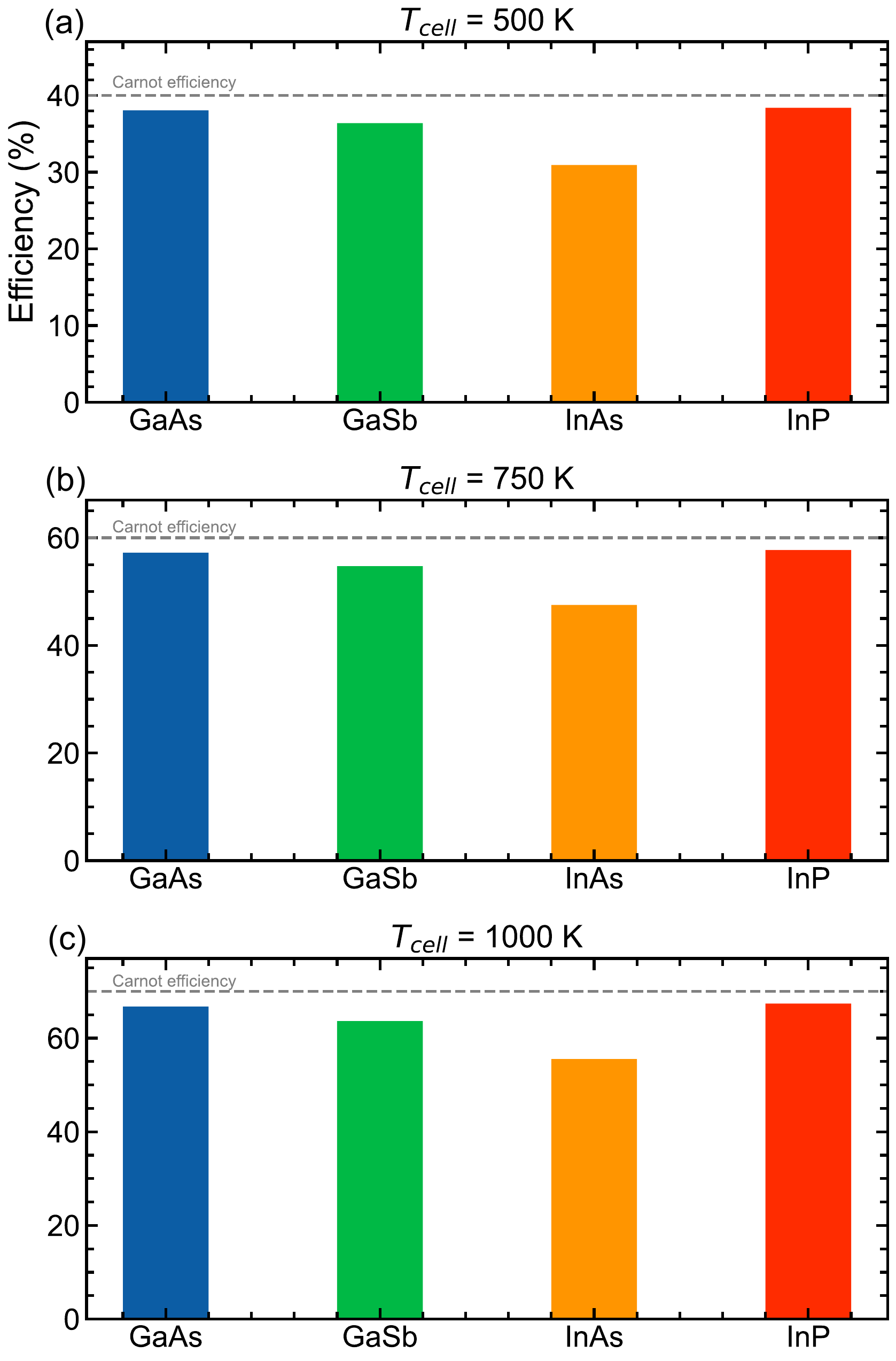}
        \caption{Comparison of thermoradiative cell efficiency with respect to different materials (GaAs, GaSb, InAs, and InP) at various cell temperatures: (a) $500$, (b) $750$, and (c) $1000$ K). The ambient temperature is set constant at $300$ K. The Carnot efficiency, represented by grey dashed lines, indicates the maximum efficiency that a TR cell can achieve.}
        \label{fig:comparison-eff-carnot}
    \end{figure} 
    
After obtaining the electronic band structure of each material, we calculate the complex dielectric function $\epsilon(\omega)$ within the independent particle and dipole approximation~\cite{Giustino2014}.  This calculation yields the real part $\epsilon_{1}(\omega)$ and imaginary part $\epsilon_{2}(\omega)$ of the dielectric function, where $\omega$ represents the photon frequency.  Using $\epsilon_{1}(\omega)$ and $\epsilon_{2}(\omega)$, we compute the frequency-dependent refractive index and extinction coefficient, which can be expressed as~\cite{Fox2010}
    \begin{align}
       n(\omega) = \sqrt{\frac{\left(\epsilon^{2}_{1}+\epsilon^{2}_{2}\right)^{1/2}_{~}+\epsilon_{1}}{2}},
    \end{align}
and
    \begin{align}
    \kappa(\omega) = \sqrt{\frac{\left(\epsilon^{2}_{1}+\epsilon^{2}_{2}\right)^{1/2}_{~}-\epsilon_{1}}{2}},
    \end{align}    
respectively.  We can further calculate the absorption coefficient by applying the relationship $\alpha(\omega) = 2\kappa\omega /c$\cite{Fox2010}.  The refractive index and extinction coefficient as functions of photon energy for four semiconducting materials (GaAs, GaSb, InAs, and InP) are shown in Fig~\ref{fig:optical}.  Utilizing these data, we can integrate them in equation~\eqref{eq:2}, and equations~\eqref{eq:4} --\eqref{eq:5} to determine the photon and radiative heat flux.

    \begin{table*}
        \centering        
        \begin{tabular}{ccccccc}
            \hline
            \multirow{2}{*}{Material} & \multirow{2}{*}{$T_{\mathrm{cell}}$ (K)} & \multirow{2}{3cm}{Maximum Power Density ($\text{W}/\text{m}^2$)} & \multicolumn{4}{c}{Efficiency (\%)} \\ \cline{4-7} 
            & & & w/o losses & w/ sub-bandgap loss & w/ heat loss & w/ all losses \\ \hline
            GaAs & 500 & $8.58 \times 10^{-6}$ & 38.06 & $2.99 \times 10^{-5}$ & $2.15 \times 10^{-6}$ & $2.00\times 10^{-6}$ \\ 
             & 750 & 0.0625 & 57.23 & 0.0174 & 0.00694 & 0.00496 \\ 
             & 1000 & 6.844 & 66.75 & 0.306 & 0.4573 & 0.188 \\ \hline
            GaSb & 500 & 0.0908 & 36.38 & 0.1108 & 0.0226 & 0.0188 \\ 
             & 750 & 19.862 & 54.72 & 1.755 & 1.7929 & 0.9738 \\ 
             & 1000 & 391.031 & 63.63 & 5.5718 & 9.9204 & 4.5831 \\ \hline
            InAs & 500 & 28.163 & 30.94 & 11.168 & 4.6355 & 4.0467 \\ 
             & 750 & 599.752 & 47.53 & 23.50 & 16.9234 & 14.965 \\ 
             & 1000 & 3825.722 & 55.53 & 32.486 & 27.6341 & 24.990 \\ \hline
            InP & 500 & $8.59 \times 10^{-9}$ & 38.38 & $4.29 \times 10^{-7}$ & $2.15 \times 10^{-8}$ & $2.04 \times 10^{-8}$ \\ 
             & 750 & 0.0035 & 57.73 & 0.00144 & 0.000394 & 0.00031 \\ 
             & 1000 & 0.9186 & 67.38 & 0.0618 & 0.0649 & 0.0318 \\ \hline
        \end{tabular}
        \caption{Quantifying the loss mechanism and its effect on the TR cell efficiency with different materials (GaAs, GaSb, InAs, and InP) at various cell temperatures (500~K, 700~K, and 1000~K) and ambient temperature of 300 K.}
        \label{table:comparison-result}
    \end{table*}

    \begin{figure}
        \centering
        \includegraphics[clip,width=8cm]{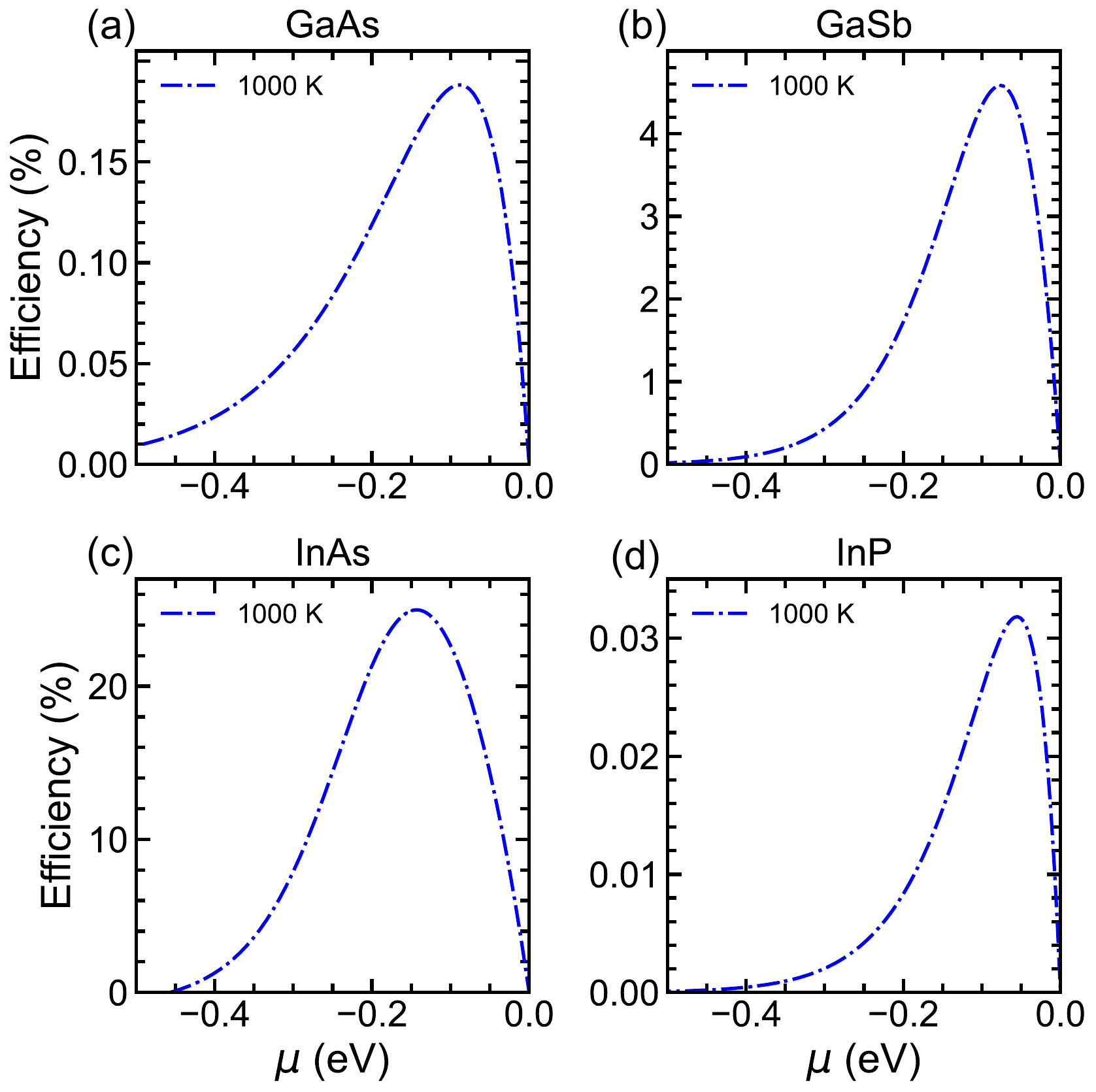}
        \caption{Efficiency of thermoradiative (TR) cells as a function of chemical potential for various materials, including (a) GaAs, (b) GaSb, (c) InAs, and (d) InP. The calculations were conducted at different cell temperatures of $1000$ K and an ambient temperature of $300$ K, taking into account sub-bandgap and heat losses.}
        \label{fig:efficiency-all-losses}
    \end{figure}
    
Figures~\ref{fig:power-density} and ~\ref{fig:efficiencywolosses} show the power density and efficiency of a thermoradiative cell without losses as a function of chemical potential at various cell temperatures $T_c$ ($500$, $750$, and $1000$ K) and ambient temperature $T_a = 300$ K. According to the results in Table 2, GaSb has the highest power density, reaching a maximum value of $391.031 \, \text{W}/\text{m}^2$ at a cell temperature of $1000$ K. Conversely, InP has the lowest power density, achieving a maximum value of $8.586\times10^{-8} \, \text{W}/\text{m}^2$ at a cell temperature of $500$ K.  Figure~\ref{fig:efficiency-all-losses} shows that maximum efficiency without losses is achieved when the materials have an energy gap larger than $k_B T$, approaching the Carnot efficiency, $\eta_{\text{Carnot}} = 1 - T_a/T_c$~\cite{Strandberg2015}. For example, when $T_c = 1000$ K and $T_{a} = 300$ K, the Carnot efficiency is about $70\%$. In the ideal condition, GaAs and InP, which have large band gaps, exhibit TR cell efficiencies of about $66.75\%$ and $67.38\%$, respectively, which are close to the Carnot efficiency. 

It is important to note that the Carnot efficiency represents the upper limit of any TR cell efficiency with no loss.  Interestingly, InAs can achieve higher efficiency than the other materials when losses are introduced (sub-bandgap and heat losses), reaching an efficiency of about $24.990 \, \%$ at a cell temperature of 1000 K,  as shown in Figure 8. The physical origin behind these results is a combination of the effects from, mainly, the bandgap energy, carrier mobility, and effective mass of carriers, among others. The band gap is, in particular, the most important factor when taking the sub-bandgap loss into account. Increasing the band gap of the material will increase the TR efficiency under the ideal condition. However, the larger the bandgap energy is, the larger the sub-bandgap loss will be. Therefore, InAs with the smallest bandgap energy among the materials considered in this work is the least affected by the sub-bandgap contribution and is thus more resistant to the losses. 

To better comprehend the interplay between power density, efficiency, and material properties, it is essential to investigate the impact of the energy gap on TR cell performance.  Figures~\ref{fig:power-density} and ~\ref{fig:efficiencywolosses} demonstrate the critical role the energy gap plays in determining efficiency and power density, especially under varying temperature conditions. The material's band gap is vital for converting thermal radiation into electricity, a process achieved through the direct recombination of electrons and holes without momentum change. This characteristic is typical of materials with a direct band gap, which greatly influences TR cell performance optimization.

\begin{figure}[tb]
        \centering
        \includegraphics[clip, width=8cm]{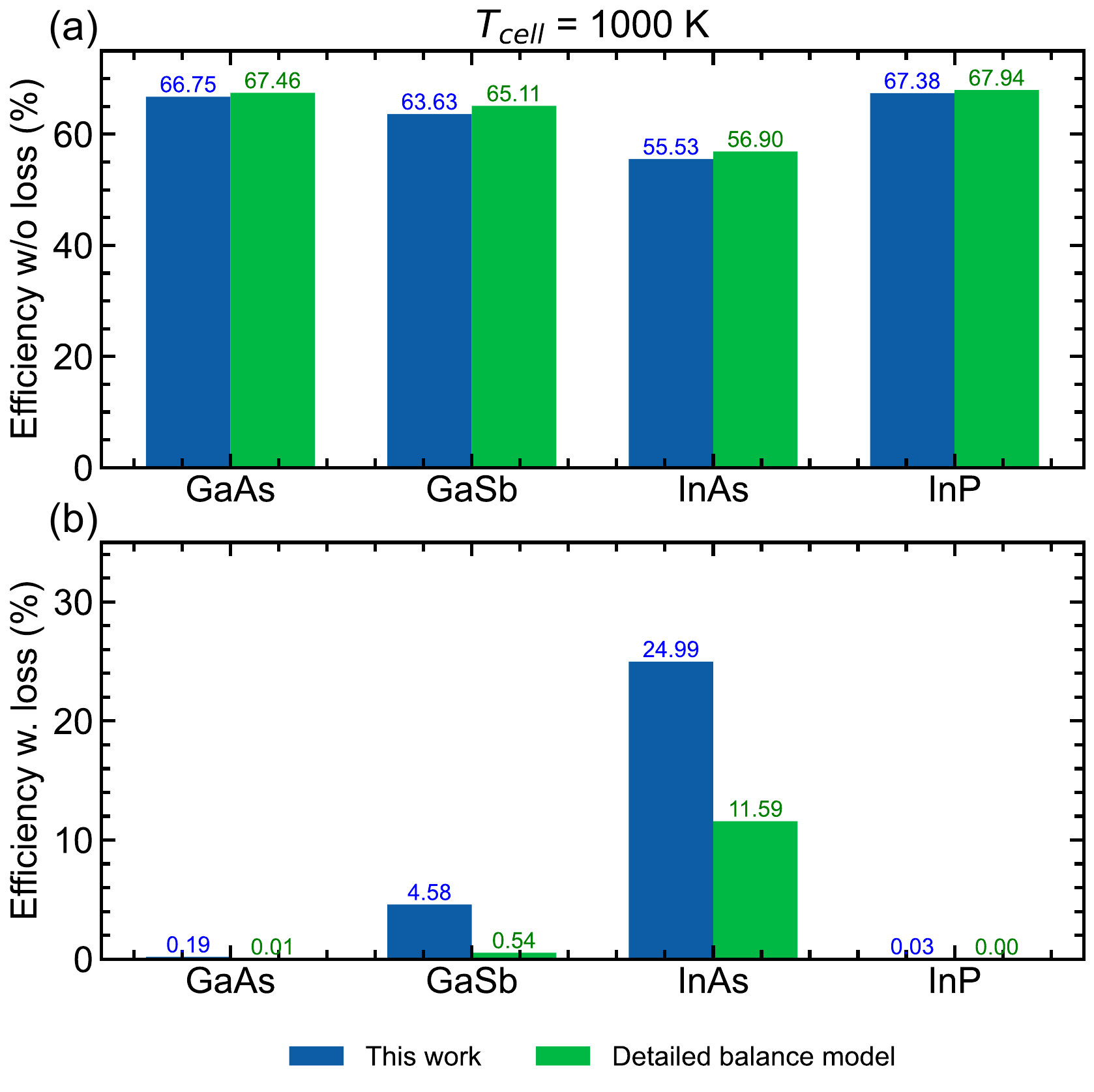}
        \caption{Comparison of TR cell efficiency (a) without losses and (b) with the inclusion of sub-bandgap and heat losses for materials including GaAs, GaSb, InAs, and InP at the cell temperature of $1000$ K.}
        \label{fig:comparison-efficiency}
    \end{figure}

To examine the influence of absorptivity on TR cell efficiency, we compare our model to the detailed-balance model by Shockley and Queisser~\cite{Shockley1961}, setting $a=0$ for photon energy below the energy gap and $a=1$ above it~\cite{Wurfel1982,Luque1997,Boriskina2016}. Figure~\ref{fig:comparison-efficiency}(a) shows a small discrepancy in efficiency between our model and the detailed-balance model without losses. However, when sub-bandgap and heat losses are introduced, particularly for InAs, our model incorporating absorptivity based on material optical properties yields significantly higher efficiency than the detailed-balance model. The absorptivity correction is expected to be crucial for accurately predicting TR cell efficiency. Notably, while sub-bandgap and heat loss does not affect power density, it contributes to efficiency due to the increased heat flow density $Q_{\mathrm{in}}$ represented by ($q_{C} + q^{\mathrm{em}}_{\hbar \omega < \hbar \omega_{g}} - q^{\mathrm{abs}}_{\hbar \omega < \hbar \omega_{g}}$)~\cite{Strandberg2015,Hsu2016,Zhang2017,Zhang2019}.
    
According to the findings presented in Table~\ref{table:comparison-result}, materials exhibiting a reduced bandgap have demonstrated suppression of sub-bandgap losses within the bulk system. Consequently, bandgap engineering emerges as a viable strategy for addressing this issue. One method to mitigate sub-bandgap losses involves the utilization of intermediate band thermoradiative (TR) cells (IBTRCs)~\cite{Ye2018}, which incorporate an intermediate band within the main absorber's bandgap. Alternatively, employing low-dimensional materials such as two-dimensional, one-dimensional, or nanostructured systems may also mitigate sub-bandgap losses~\cite{Feng2022,Chaves2020}. These materials exhibit tunable bandgaps that can be engineered to correspond with the thermal radiation spectrum of a specific emitter.  

\section{Conclusions and Outlook}
We have conducted simulations to evaluate TR cell performance by examining the band gaps and optical properties of direct-gap semiconducting materials, such as GaAs, GaSb, InAs, and InP. Our results indicate that a larger band gap increases efficiency but decreases power density under ideal conditions. On the other hand, as the bandgap value decreases, the presence of sub-bandgap and heat losses still can preserve appreciable efficiency values.  We compare the detailed-balance model with our absorptivity-based model, which is more suitable for accurately predicting TR cell efficiency when the sub-bandgap and heat losses are considered. It is important to note that these losses decrease efficiency by increasing heat flow density, not power density.

By analyzing TR cell efficiency through simulations, we explore the potential of bandgap engineering to optimize performance.  We expect that suppressing the sub-bandgap loss with low-dimensional materials tailored to match an emitter's thermal radiation spectrum can be achieved through bandgap engineering. This study provides valuable insights into TR cell efficiency, potentially contributing to the advancement of sustainable energy technology.  We note that we still could not compare our simulations directly with the available experimental results of the III-V group semiconductor because of the limited experimental data. Therefore, we instead expect that our work would trigger further experimental investigations based on the predicted efficiencies from our simulation for the group III-V semiconductors.

\section*{Acknowledgments}
M.Y.H. is supported by the "Degree-by-Research" program from the National Research and Innovation Agency (BRIN).  We also acknowledge Mahameru BRIN for its high-performance computing facilities.

\section*{References}
\bibliographystyle{iopart-num}
\bibliography{references.bib}

\end{document}